\documentclass[prd,preprintnumbers,showpacs,12pt]{revtex4}
\usepackage{epsfig}
\usepackage{graphicx}
\usepackage {axodraw}

\usepackage{bm}
\usepackage{amsmath}
\usepackage{amssymb}
\usepackage{amscd}
\usepackage{latexsym}

\newcommand{\be}{\begin{equation}}

\newcommand{\ee}{\end{equation}}
\newcommand{\bea}{\begin{eqnarray}}
\newcommand{\eea}{\end{eqnarray}}

\newcommand{\nn}{\nonumber}

\def\Black{}
 \def\AliasBlue{}
 \def\Blue{}
 \def\Brown{}

\begin{document}

\newcommand{\bra}[1]{\langle #1|}
\newcommand{\ket}[1]{|#1\rangle}
\newcommand{\braket}[2]{\langle #1|#2\rangle}
\newcommand{\tr}{\textrm{Tr}}
\newcommand{\lag}{\mathcal{L}}
\newcommand{\mbf}[1]{\mathbf{#1}}
\newcommand{\desl}{\slashed{\partial}}
\newcommand{\Desl}{\slashed{D}}

\renewcommand{\bottomfraction}{0.7}
\newcommand{\epsi}{\varepsilon}

\newcommand{\nl}{\nonumber \\}
\newcommand{\tc}[1]{\textcolor{#1}}
\newcommand{\sla}{\not \!}
\newcommand{\spinor}[1]{\left< #1 \right>}
\newcommand{\cspinor}[1]{\left< #1 \right>^*}
\newcommand{\Log}[1]{\log \left( #1\right) }
\newcommand{\Logq}[1]{\log^2 \left( #1\right) }
\newcommand{\mr}[1]{\mathrm{#1}}
\newcommand{\cw}{c_\mathrm{w}}
\newcommand{\sw}{s_\mathrm{w}}
\newcommand{\ct}{c_\theta}
\newcommand{\st}{s_\theta}
\newcommand{\gt}{{\tilde g}}
\newcommand{\gtp}{{{\tilde g}^\prime}}
\renewcommand{\i}{\mathrm{i}}
\renewcommand{\Re}{\mathrm{Re}}
\newcommand{\yText}[3]{\rText(#1,#2)[][l]{#3}}
\newcommand{\xText}[3]{\put(#1,#2){#3}}


\def\to{\rightarrow}
\def\ptl{\partial}
\def\beq{\begin{equation}}
\def\eeq{\end{equation}}
\def\bea{\begin{eqnarray}}
\def\eea{\end{eqnarray}}
\def\nn{\nonumber}
\def\half{{1\over 2}}
\def\rhalf{{1\over \sqrt 2}}
\def\calo{{\cal O}}
\def\call{{\cal L}}
\def\calm{{\cal M}}
\def\del{\delta}
\def\eps{\epsilon}
\def\lam{\lambda}

\def\anti{\overline}
\def\delfac{\sqrt{{2(\del-1)\over 3(\del+2)}}}
\def\heff{h'}
\def\square{\boxxit{0.4pt}{\fillboxx{7pt}{7pt}}\hspace*{1pt}}
    \def\boxxit#1#2{\vbox{\hrule height #1 \hbox {\vrule width #1
             \vbox{#2}\vrule width #1 }\hrule height #1 } }
    \def\fillboxx#1#2{\hbox to #1{\vbox to #2{\vfil}\hfil}   }

\def\braket#1#2{\langle #1| #2\rangle}
\def\gev{~{\rm GeV}}
\def\gam{\gamma}
\def\sn{s_{\vec n}}
\def\sm{s_{\vec m}}
\def\mm{m_{\vec m}}
\def\mn{m_{\vec n}}
\def\mh{m_h}
\def\sumn{\sum_{\vec n>0}}
\def\summ{\sum_{\vec m>0}}
\def\vl{\vec l}
\def\vk{\vec k}
\def\ml{m_{\vl}}
\def\mk{m_{\vk}}
\def\gp{g'}
\def\gt{\tilde g}
\def\hw{{\hat W}}
\def\hz{{\hat Z}}
\def\ha{{\hat A}}

\def\yy{{\cal Y}_\mu}
\def\yyt{{\tilde{\cal Y}}_\mu}
\def\lq{\left [}
\def\rq{\right ]}
\def\dmu{\partial_\mu}
\def\dnu{\partial_\nu}
\def\dmus{\partial^\mu}
\def\dnus{\partial^\nu}
\def\gp{g'}
\def\gpt{{\tilde g'}}
\def\ggs{\frac{g}{\gs}}
\def\eps{{\epsilon}}
\def\tr{{\rm {tr}}}
\def\V{{\bf{V}}}
\def\W{{\bf{W}}}
\def\Wt{\tilde{ {W}}}
\def\Y{{\bf{Y}}}
\def\Yt{\tilde{ {Y}}}
\def\L{{\cal L}}
\def\s{s_\theta}
\def\st{s_{\tilde\theta}}
\def\c{c_\theta}
\def\ct{c_{\tilde\theta}}
\def\gt{\tilde g}
\def\et{\tilde e}
\def\At{\tilde A}
\def\Zt{\tilde Z}
\def\Wpt{{\tilde W}^+}
\def\Wmt{{\tilde W}^-}

\newcommand{\Apt}{{\tilde A}_1^+}
\newcommand{\Bpt}{{\tilde A}_2^+}
\newcommand{\Amt}{{\tilde A}_1^-}
\newcommand{\Bmt}{{\tilde A}_2^-}
\newcommand{\Wtp}{{\tilde W}^+}
\newcommand{\Atp}{{\tilde A}_1^+}
\newcommand{\Btp}{{\tilde A}_2^+}
\newcommand{\Atm}{{\tilde A}_1^-}
\newcommand{\Btm}{{\tilde A}_2^-}
\def\mathswitchr#1{\relax\ifmmode{\mathrm{#1}}\else$\mathrm{#1}$\fi}
\newcommand{\Pe}{\mathswitchr e}
\newcommand{\Pp}{\mathswitchr {p}}
\newcommand{\PZ}{\mathswitchr Z}
\newcommand{\PW}{\mathswitchr W}
\newcommand{\PD}{\mathswitchr D}
\newcommand{\PU}{\mathswitchr U}
\newcommand{\PQ}{\mathswitchr Q}
\newcommand{\Pd}{\mathswitchr d}
\newcommand{\Pu}{\mathswitchr u}
\newcommand{\Ps}{\mathswitchr s}
\newcommand{\Pc}{\mathswitchr c}
\newcommand{\Pt}{\mathswitchr t}
\newcommand{\rd}{{\mathrm{d}}}
\newcommand{\GW}{\Gamma_{\PW}}
\newcommand{\GZ}{\Gamma_{\PZ}}
\newcommand{\GeV}{\unskip\,\mathrm{GeV}}
\newcommand{\MeV}{\unskip\,\mathrm{MeV}}
\newcommand{\TeV}{\unskip\,\mathrm{TeV}}
\newcommand{\fba}{\unskip\,\mathrm{fb}}
\newcommand{\pba}{\unskip\,\mathrm{pb}}
\newcommand{\nba}{\unskip\,\mathrm{nb}}
\newcommand{\PT}{P_{\mathrm{T}}}
\newcommand{\PTmiss}{P_{\mathrm{T}}^{\mathrm{miss}}}
\newcommand{\CM}{\mathrm{CM}}
\newcommand{\inv}{\mathrm{inv}}
\newcommand{\sig}{\mathrm{sig}}
\newcommand{\tot}{\mathrm{tot}}
\newcommand{\backg}{\mathrm{backg}}
\newcommand{\evt}{\mathrm{evt}}
\def\mathswitch#1{\relax\ifmmode#1\else$#1$\fi}
\newcommand{\M}{\mathswitch {M}}
\newcommand{\R}{\mathswitch {R}}
\newcommand{\TEV}{\mathswitch {TEV}}
\newcommand{\LHC}{\mathswitch {LHC}}
\newcommand{\MW}{\mathswitch {M_\PW}}
\newcommand{\MZ}{\mathswitch {M_\PZ}}
\newcommand{\Mt}{\mathswitch {M_\Pt}}
\newcommand{\gs}{{g''}^2}
\def\lmu{{\bf L}_\mu}
\def\rmu{{\bf R}_\mu}
\def\si{\sigma}
\def\beqar{\begin{eqnarray}}
\def\eeqar{\end{eqnarray}}
\def\refeq#1{\mbox{(\ref{#1})}}
\def\reffi#1{\mbox{Fig.~\ref{#1}}}
\def\reffis#1{\mbox{Figs.~\ref{#1}}}
\def\refta#1{\mbox{Table~\ref{#1}}}
\def\reftas#1{\mbox{Tables~\ref{#1}}}
\def\refse#1{\mbox{Sect.~\ref{#1}}}
\def\refses#1{\mbox{Sects.~\ref{#1}}}
\def\refapps#1{\mbox{Apps.~\ref{#1}}}
\def\refapp#1{\mbox{App.~\ref{#1}}}
\def\citere#1{\mbox{Ref.~\cite{#1}}}
\def\citeres#1{\mbox{Refs.~\cite{#1}}}

\def\Black{}
 \def\AliasBlue{}
 \def\Blue{}
 \def\Brown{}


\title{Discovery and Mass Spectroscopy via
Mixed Di-Boson Production \\at the LHC
in a 4-Site Model with a Composite Higgs Boson}

\date{\today}
 \author{Luca Fedeli$^*$ and Stefano Moretti}%
 \email{l.fedeli@soton.ac.uk; s.moretti@soton.ac.uk}
 \affiliation{School of Physics and Astronomy, University of
 Southampton, Highfield,
 Southampton SO17 1BJ, UK}%
\author{Stefania De Curtis}%
 \email{decurtis@fi.infn.it}
 \affiliation{Istituto Nazionale di Fisica Nucleare, Sezione di Firenze,  Via G. Sansone 1, 50019 Sesto Fiorentino, Italy}%
 \author{Daniele Dominici}%
 \email{dominici@fi.infn.it}
 \affiliation{Universit\`a degli Studi di Firenze, Dip. di
 Fisica e Astronomia, \\
 and Istituto Nazionale di Fisica Nucleare, Sezione di Firenze,  
Via G. Sansone 1, 50019 Sesto Fiorentino, Italy}%

\begin{abstract}
\noindent
\small
We investigate the scope of the Large Hadron Collider (LHC)
in probing the parameter space of a 4-Site (4S) model supplemented
by one composite Higgs state. We concentrate on the yield of `mixed' di-boson production giving three leptons and missing (transverse)
energy in the final state, i.e., events induced via the sub-process $q\bar q'\to  l^+\nu_l~ l^{'+}l^{'-}$ +  ${\rm{c.c.}}$, 
where $q,q'$ are quarks and $l,l'=e,\mu$ in all possible combinations, signature
which enables the production in the intermediate
steps of all additional -- with respect to the Standard Model (SM) -- neutral and charged gauge bosons belonging to the spectrum 
of this scenario, all of which in resonant topologies.
We not only find this channel to be accessible over the background  but also show that, 
after a dedicated cut-based analysis, kinematical reconstruction of all such resonances is always possible. 
We finally compare the yield of the mixed di-boson mode to that of the `charged' di-boson one and find that the former is of lesser scope than the latter at
smaller gauge boson masses but it overcomes it at higher values, thereby revealing itself a viable search mode covering new regions of parameter space.
\end{abstract}

\vspace*{-1.0truecm}
\maketitle
\section{Introduction}
\label{sec:intro}
A strongly broken Electro-Weak (EW) sector is expected to contain a variety of bound states including particles of 
spin 0, spin 1/2 as well as
spin 1. Its phenomenology below the scale of the  new interactions responsible for EW Symmetry Breaking (EWSB)
and producing such new resonances can be studied, amongst others, in simple four-dimensional models, like the 4-Site 
(4S) one of \cite{Accomando:2008jh}.

In its original formulation, the 4S model describes in an effective way the interactions of extra spin 1 resonances (four in total) as gauge fields of a $SU(2)\otimes SU(2)$ extra gauge group. They can be thought of as the first
Kaluza-Klein (KK) excitations emerging from a five-dimensional formulation, and, due to the Anti-de Sitter/Conformal Field Theory
(AdS/CFT) correspondence, they are composite states of a strong dynamics also responsible for the breaking of the EW symmetry.  In this note we consider a 4S model
supplemented by the inclusion of a new scalar field, singlet under the above gauge group, in order to reproduce, in our effective description, the scalar particle recently detected by the ATLAS and CMS experiments at the Large Hadron Collider (LHC) \cite{:2012gk,:2012gu}. The model has been introduced in \cite{Accomando:2012yg}, to which
we refer the reader for technical details. In such a scenario, the couplings of our composite Higgs-like scalar particle to the 
Standard Model (SM) and extra gauge bosons are free parameters for which we have derived bounds due to the EW Precision Tests (EWPTs) and the present measurements at the 
Tevatron and LHC, via Drell-Yan (DY) processes, as well as theoretical constraints enforced by unitarity requirements.

It is the purpose of this paper to investigate, in the context of such a 4S model with one composite Higgs 
state (compatible with Tevatron and LHC data), the phenomenology of `mixed' di-boson production at the LHC, yielding three leptons and missing transverse energy, i.e., the process
\begin{equation}\label{eq:process}
pp(q\bar q')\to W^\pm Z\to l^\pm  l^{'+}l^{'-}  E_{T}^{\rm miss},
\end{equation}
wherein the symbols $W^\pm$ and $Z$ refer to any possible charged and neutral, respectively, spin 1 massive gauge bosons present in the model,
$q$ and $q'$ are the quarks found inside the proton
whilst $l$ and $l'$ refer to either an electron $e$ or a muon $\mu$, the latter two in all possible combinations of (identical and
different) such flavors. 

In the attempt to extract the signal in (\ref{eq:process}), we will focus our attention onto all energy and luminosity stages covered already or 
still foreseen for
the CERN machine. Ultimately, we will want to contrast the detection
potential at the LHC of mixed di-boson production  with that of both DY,
\begin{equation}\label{eq:processDY}
q\bar q'\to W^\pm\to l^\pm  E_{T}^{\rm miss},
 \qquad\qquad\qquad q\bar q\to Z\to l^+l^-,
\end{equation}
 and `charged' di-boson events,
\begin{equation}\label{eq:processWW}
pp(q\bar q)\to W^+W^-\to e^\pm\mu^\mp E_{T}^{\rm miss},
\end{equation} 
building on
previous studies of ours. Further, we will show that the presence of only one source of missing transverse energy
(the neutrino $\nu_\ell$) in the case of mixed di-boson production enables one to extract
simultaneously, upon reconstructing its longitudinal momentum, all masses of the new gauge bosons present in the model. 

The plan of the paper is as follows. The next section will describe the model at hand.
The following one, after describing the calculation, will be devoted to study the production and decay dynamics of process
(\ref{eq:process}), eventually extracting from it exclusion and discovery limits over the accessible parameter space and contrasting these to the yield of processes (\ref{eq:processDY}) and (\ref{eq:processWW}).
Then, in the same section, we will also illustrate how to extract the mass spectrum of the 4S model. A final section will be devoted to
summarize our work and conclude. {In Appendix~\ref{app1}, owing to their importance for this analysis, we will list  approximated (yet accurate) analytical expressions for the three gauge boson vertices.}

\section{The 4S Model with a Singlet Composite Scalar  Higgs State}

The 4S model is a moose model based on a $SU(2)_L \otimes  SU(2)_1\otimes SU(2)_2 \otimes U(1)_Y$ gauge symmetry and contains three non-linear $\sigma$-model fields interacting with the gauge fields, which trigger spontaneous 
EWSB. Its construction is presented in \cite {Accomando:2008jh,Accomando:2012yg} while some of its phenomenological consequences are analysed in 
\cite{Accomando:2008dm,Accomando:2010fz,Accomando:2010ir,Accomando:2011vt,Accomando:2011xi,Accomando:2011eu}.
In the unitary gauge, the spectrum of the 4S model incorporates (other than all the SM particles, including the Higgs boson, which is however not responsible for EWSB) two new triplets of gauge 
bosons, which acquire mass through the same non-linear symmetry breaking 
mechanism giving mass to the SM gauge bosons. We denote with 
$W_{i\mu}^\pm$ and $Z_{i\mu}$ ($i = 1, 2$) the four charged and two neutral 
heavy resonances, respectively, appearing as a consequence of the gauge group extension, and 
with $W^\pm_\mu$, $Z_\mu$ and $A_\mu$ the SM gauge bosons. Owing to its gauge 
structure, the 4S
model a priori contains seven free parameters: the $SU(2)_L\otimes U(1)_Y$ 
gauge couplings, $\tilde g$ and $\tilde g'$, the extra $SU(2)_{1,2}$ gauge 
couplings that we assume to be equal, $g_2= g_1$, due to a $LR$ symmetry
\cite{Accomando:2008jh}, the bare 
masses of lighter ($W_1^\pm, Z_1$) and heavier ($W_2^\pm, Z_2$) gauge boson 
triplets, $M_{1,2}$, and their bare direct couplings to SM fermions, $b_{1,2}$,
as described in \cite{Accomando:2008jh}. 
However, their number can be reduced to four, by fixing the gauge couplings 
$\tilde g,\tilde g', g_1$ in terms of the three SM input parameters 
$\alpha, G_F, M_Z$, which denote the Electro-Magnetic (EM) and Fermi constants plus the $Z$ boson mass, 
respectively. As a result, the parameter space is completely defined by  
four independent free parameters, which one can choose to be: $M_1$, $z$, $b_1$ and 
$b_2$, where $z=M_1/M_2$ is the ratio between the bare masses. In terms of 
these four parameters, physical masses and couplings of the extra gauge bosons 
to ordinary matter can be obtained via a complete numerical algorithm 
\cite{Accomando:2011vt}. In the 
following, we choose to describe the full parameter space via the physical 
observables: other than $z$ (which, as shown in 
\cite{Accomando:2011vt}, is a good approximation of the ratio between physical 
masses $M_{W_1}/M_{W_2}$ or $M_{Z_1}/M_{Z_2}$), we take  $M_{1}, a_{W_1}$ and $a_{W_2}$ which denote the mass 
of the lighter 
extra  gauge boson and the couplings of the lighter and heavier extra 
 gauge bosons to ordinary matter, respectively. 

We refer the reader to Refs.~\cite{Accomando:2008jh,Accomando:2012yg} for all details about theoretical and experimental bounds, mass spectrum and new resonance properties of the model being studied here. 

\section{Results}
\label{sec:results}
\subsection{Calculation}
\begin{figure}[!tbp]
\begin{center}
\hspace*{-1.0cm}
\includegraphics[scale=0.75]{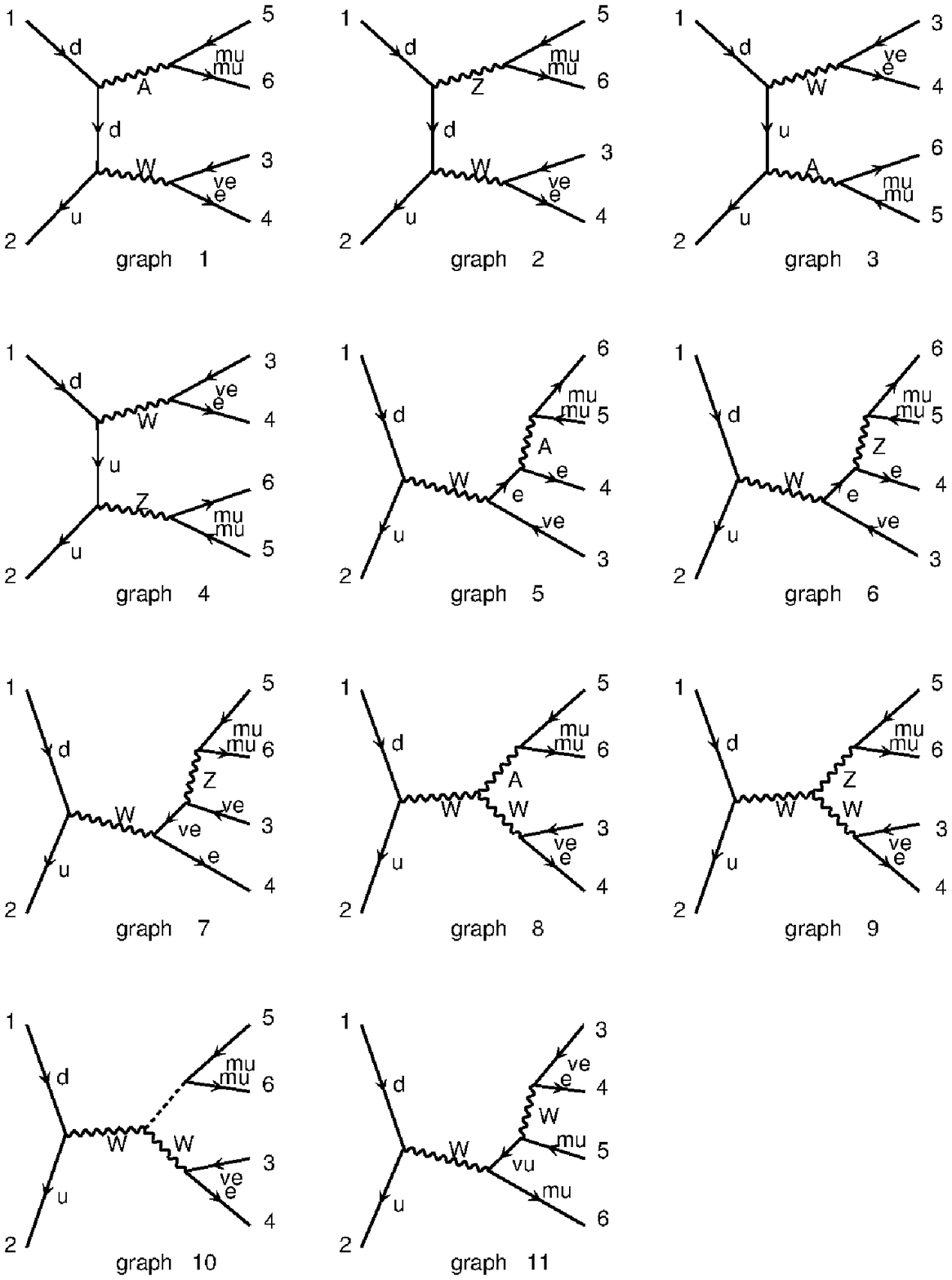}
\end{center}
\vspace{-5.5cm}
\caption{Topologies of Feynman diagrams for the process in eq.~(\ref{eq:process}),
in the case $l\ne l'$ (in this case, e.g., $l=e$ and $l'=\mu$). For the case $l=l'$, one ought
to swap helicities/momenta of two identical particles and antisymmetrize the diagrams (see
text). Here the labels $Z$ and $W$ refer to all possible {massive} gauge boson states of the model, neutral and charged,
respectively. {The dashed line in graph 10 represents the scalar Higgs boson (this diagram is negligible though due to the smallness of the Yukawa couplings involved)}.} 
\protect{\label{fig:diagrams}}
\end{figure}

The codes exploited for our study of the LHC signatures
are based on helicity amplitudes, defined through the HELAS subroutines~\cite{Murayama:1992gi}
and assembled by means of 
MadGraph~\cite{Stelzer:1994ta}. 
VEGAS~\cite{Lepage:1977sw} was used for the multi-dimensional numerical integrations.
The Matrix Elements (ME) accounts for all off-shellness effects
of the particles involved and was constructed starting from the topologies in Fig.~\ref{fig:diagrams}, 
wherein (as intimated) the labels $Z$ and $W$ refer to any possible
combination of gauge bosons in our model. 
The Parton Distribution Functions (PDFs) used 
were CTEQ5L~\cite{Lai:1999wy}, with factorization/renormalization
scale set to $Q=\mu=\sqrt{\hat{s}}$. Initial state quarks have been taken
as massless, just like, unless otherwise stated, the final state leptons and neutrino. 

\subsection{Event Selection}
To calculate the Cross-Section (CS) at the LHC for the mixed di-boson channel we start considering  general purpose cuts, which we call  {\it Cc} cuts. 
The ensuing selection is designed in a such way to have three leptons in the final state, to produce
a large signal CS, thereby usable for search purposes
at low luminosities, and to give some hint about the typical mass scales of the new 
gauge bosons. This set of cuts is designed in the same spirit as those 
previously used to study processes (\ref{eq:processDY}) \cite{Accomando:2010ir,Accomando:2011xi}
and (\ref{eq:processWW}) \cite{Accomando:2012yg}, essentially based on detector acceptance.

However, as intimated in the Introduction, the presence of only one neutrino in the final state of process (\ref{eq:process}) enables
us to fully reconstruct the final state kinematics (see later on for technical details). This, as we will show,
affords us then with the possibility of not only improving the discovery power of mixed di-boson production but also 
extracting the new gauge boson resonances. However,
in order to do so, we 
will need to introduce a second, much tighter set of additional cuts, denoted by {\it C2},
which will make use of the partial knowledge acquired through the previous, looser selection
of the mass scales of the new 4S gauge boson resonances, specifically, of the lightest ones.
This set of cuts will then 
be used to both compare the exclusion/discovery potential of the LHC among the
three channels (\ref{eq:process})--(\ref{eq:processWW}) and to enable one to observe all six gauge
boson resonances (SM and 4S ones). The drawback is that the {\it C2} sample yields a smaller CS
than the {\it Cc} one, so that higher luminosities are required.
  
Before illustrating the results of the analysis, we ought to define the
observables that have been used to characterize this process. Here and in the following, we refer to the (visible) particles trough their numerical labels $i,j=4,5,6$, as listed in Fig.~\ref{fig:diagrams}.\\
\begin{itemize}
\item $P^T_{i[j]}=\sqrt{(P_i^x[+P_j^x])^2+(P_i^y[+P_j^y])^2}$ is the transverse momentum of a particle $i$[of a pair of particles $ij$]
\item $p^T_M={\rm max}(P^T_4,P^T_5,P^T_6)$ is the maximum amongst the transverse momenta of the particles  
\item $P^T_M={\rm max}(P^T_{45},P^T_{46},P^T_{56})$  is the maximum amongst the transverse momenta of all possible pairs of particles
\item $M_{ij[k]}=\sqrt{(P_i+P_j[+P_k])^\mu(P_i+P_j[+P_k])_\mu}$ is the invariant mass of a pair[tern] of particles 
\item $M^T_{ij[k]}=\sqrt{(P_i^0+P_j^0[+P_k^0])^2-(P_i^x+P_j^x[+P_k^x])^2-(P_i^y+P_j^y[+P_k^y])^2}$ 
is the transverse mass of a pair[tern] of particles 
\item $\eta_i=-\log\tan\frac{\theta_i}2$ is the pseudo-rapidity of a particle
\item $\theta_{i[j]}$ is the angle between the beam axis and a particle[between two particles]
\item $\cos\phi_{ij}=\frac{P_i^xP_j^x+P_i^yP_j^y}{P^T_i P_j^T}$ is the cosine of the relative (azimuthal) angle between two particles in the plane transverse to the beam 
\item $E_{T}^{\rm miss}=\sqrt{(P_4^x+P_5^x+P_6^x)^2+(P_4^y+P_5^y+P_6^y)^2}$ is the missing transverse energy (due to the neutrino escaping detection) 
\end{itemize}
Here, $P_i$ ($i=1, ... 6$) are the momenta of the external particles entering process (\ref{eq:process}).
In the following we list the two sets of cuts used, as detailed above.\\
\noindent $\bullet$
{\it Cc}: 
$P^T_{i}>20$~GeV, 
$M_{ij}>20$~GeV,
$E_{T}^{\rm miss}>50$~GeV,
$|\eta_i|<2$,
cos$\theta_{ij}<0.9$;\\ 
\noindent $\bullet$ {\it C2}: {\it Cc} plus
$M_{456}>0.9 ~M_{W_1}$, 
$P^T_{ij}>150$~GeV, 
$p^T_M>150$~GeV, 
$\cos\phi_{56}<-0.5$.

Before proceeding with the kinematical analysis though, a couple of subtle points should be noted.
The signatures we are considering are four, differing only for the flavor of the final state leptons, as we include the charge conjugation directly in the Monte Carlo (MC) code. In particular, the fermion flavors may be the following: ($3=e$, $4=5=\mu$), 
($3=\mu$, $4=5=e$),   ($3=4=5=e$) and ($3=4=5=\mu$). We refer to the first two cases (which give identical rates) as the electron-muon ($e\mu$) combination and for 
the last two (also identical) as the three lepton ($3l$) one. For the $e\mu$ case, in order to obtain the total CS, one straightforwardly compute the diagrams in Fig.~\ref{fig:diagrams}. In contrast,
for the $3l$ case, one must take into account the fact that there are two identical leptons in the final state
(the third charged lepton differ from the other two for its electric charge), so we have to consider an antisymmetric final state. This duplicates the number of diagrams to compute by 
exchanging the helicities ($h_{4,6}$) and momenta ($P_{4,6}$) of the two identical leptons, so that, according to Pauli's statistic, one obtains:
\be
\mathcal{M}_{3l}(...,P_4,h_4,...,P_6,h_6)=\frac 1 {\sqrt{2}}\left(\mathcal{M}_{e\mu}(...,P_4,h_4,...,P_6,h_6)-
\mathcal{M}_{e\mu}(...,P_6,h_6,...,P_4,h_4)\right).
\ee
Computing the total CS for both $e\mu$ and $3l$ we found  a small variation 
($\sigma_{3l}/\sigma_{e\mu}= 0.85-0.90$) 
for both the SM and the full 4S. Further, we have
verified that, after either set of cuts is implemented, all of the differential distributions are
essentially identical over most of the kinematic ranges considered.

\begin{figure}[!t]
\begin{center}
\unitlength1.0cm
\begin{picture}(7,4)
\put(-1.05,-4){\epsfig{file=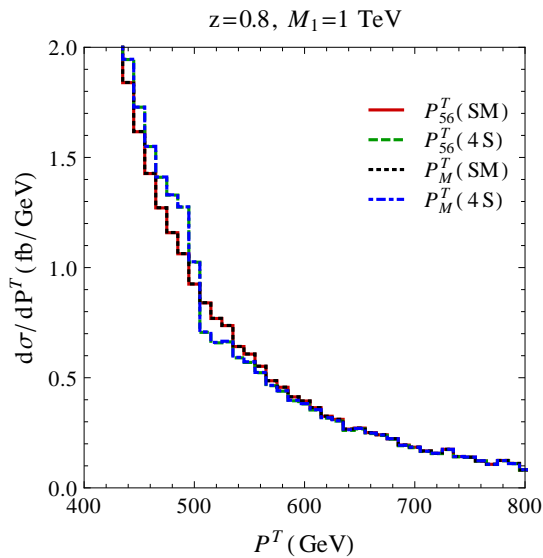,width=7.5cm}}
\end{picture}
\end{center}
\vskip 3.5cm
\caption{Differential distributions for $P_{56}^T$ and $P^T_M$ for the SM and in a benchmark point of the 4S model (in particular, we choose $z=0.8$, $M_1$=1~TeV
and $a_{W_1}=0.13$) after the {\it Cc} cuts. Notice that, on purpose, the lines are practically indistinguishable.
}
\label{fig:emu}
\end{figure}

Another important point in our analysis is the capability to identify the two leptons coming from the neutral particle ($A,Z,Z_1,Z_{2}$).
In the $e\mu$ case this is trivial, since the couple of same flavor leptons comes for sure 
from a neutral gauge boson. In the $3l$ 
scenario this is in principle ambiguous, as we cannot distinguish between the two same charge leptons (there are two identical
particles in the final state, so it is clear that they are indistinguishable), i.e., the one coming from 
the $W,W_1,W_2$ and the other coming from the $A,Z,Z_1,Z_{2}$. However if we consider $P_{45}^T$ and $P_{56}^T$ (the couple 4,6 is useless, due to the fact that 
particles 4 and 6 have the same charge and cannot come from a neutral boson) we get that almost in the 100\% of the cases 
$P^T_M=P_{56}^T$ (in fact, this is false only in 0.8\% of the cases):
Fig.~\ref{fig:emu} illustrates this.  Therefore, we are in a position to distinguish amongst leptons
also in the case of identical flavors and entitled to apply cuts to each of these individually.
The above statement is true after the {\it Cc} cuts, so that we can then exploit
the above identification when applying the {\it C2} cuts, i.e., those designed for mass spectrum
extraction, for which the knowledge of the final state momenta is a prerequisite. 

\subsection{Mixed Di-Boson Production and Decay in the 4S Model}
Before starting to consider the phenomenology of process (\ref{eq:process}) with three visible
leptons, let us consider its contribution to the di-lepton sample exploited in Ref.~\cite{Accomando:2012yg}, whereby the case of channel (\ref{eq:processWW}) was considered, limited to the different flavor case. Hence, for the process at hand here,
we ought 
to consider its contribution when, e.g., the muon with the same charge of the electron is outside the detector acceptance region. In Tab.~\ref{tab:WW} 
we list the results using the {\it So} cuts for process (\ref{eq:process})
and the \textit{So} cuts (defined in Ref.~\cite{Accomando:2012yg}) for channel
(\ref{eq:processWW}), for past, present and future energy configurations of the LHC. Here,
the benchmark point is defined by  $z=0.8$, $M_1=1$~TeV and $a_{W_1}=0.13$ \footnote{Notice that $z=0.8$ is the case for which processes
(\ref{eq:processDY}) and (\ref{eq:processWW}) have comparable scope at the LHC, as seen in Ref.~\cite{Accomando:2012yg}.}.
As we can see the $WZ$ contribution is absolutely negligible compared to the $WW$ one, to
both the Background ($B$), which is the  SM rate, and the Signal ($S$), defined as the difference
between the full 4S rate and the SM one \footnote{Notice that, in performing this exercise, in order to obtain a finite
result, we ought to keep a finite mass for the muon, which we take to be
$m_\mu=0.105$ GeV.}. This is a general feature of the 4S model in fact, valid over its
entire parameter space. 
\begin{table}[!b]
\begin{center}
\begin{tabular}{||c|c|c||c|c|c||c|c|c||}
\hline \hline
$\sqrt s=$~7 TeV&$WW$~(fb) & $WZ$~(fb)&$\sqrt s=$~8 TeV&$WW$~(fb) & $WZ$~(fb)&$\sqrt s=$~14 TeV&$WW$~(fb) & $WZ$~(fb) \\
\hline 
\hline
$B$&0.58&0.035&$B$&0.73&0.081&$B$&1.55&0.10\\
\hline 
$S$&2.97&0.0003&$S$&3.99&0.0004&$S$&13.4&0.0013\\
\hline\hline
\end{tabular}
\end{center}
\caption{Values of the CS for the 4S comparing processes (\ref{eq:process}), herein labeled $WZ$,  
and (\ref{eq:processWW}), herein labeled $WW$, for the case of a different flavor di-lepton signature, after the 
{\it Cc} cuts for the former and the 
{\it So} cuts (defined in  Ref.~\cite{Accomando:2012yg}) for the latter. {(In essence, here, we are quantifying
the probability that one of the two same-flavor leptons in the $WZ$ case escapes detection thereby mimicking the final
state naturally induced by the $WW$ case.)}}
\label{tab:WW}
\end{table}

The situation is rather different for the three-lepton signatures,  $e\mu$ and $3l$, that, from
now on, we will treat cumulatively. In view of the fact that we will show that process (\ref{eq:process})
has some potential at the LHC in accessing the parameter space of the 4S model, it is
worthed to review here the latter in some detail and explain how it
can influence the CS for process (\ref{eq:process}). Of the aforementioned four independent 
parameters uniquely defining the 4S model,  $z$, $M_1(\sim M_{W_1}\sim M_{Z_1})$, $a_{W_1}$ and $a_{W_2}$, we choose the first
three as input parameters and fix the other in a such way to satisfy the EWPTs (namely $\eps_{1,2,3}$). This means that, for each set of
$z$, $M_1$ and $a_{W_1}$, there is a maximal ($a_{W_2}^M$) and a minimal ($a_{W_2}^m$) value allowed for $a_{W_2}$, 
e.g., for $z=0.8$, $M_1=1$~TeV, $a_{W_1}=0.13$ we have 
$-0.185\le a_{W_2}\le-0.174$. To map the plane $(M_{W_1},a_{W_1})$ we need to make a choice to fix $a_{W_2}.$
Two possibilities are \footnote{As choice (a) was made in Refs.~\cite{Accomando:2010ir,Accomando:2011xi,Accomando:2012yg}, to which we will compare, we will maintain it here as our default, though
we will comment regarding the alternative choice.}:
\be
\label{eq:a}
(a)\quad a_{W_2}^a=\left\{\begin{array}{l}
                       a_{W_2}^M\quad \mbox{if}\quad |a_{W_2}^M|>|a_{W_2}^m|\\
                       a_{W_2}^m\quad \mbox{otherwise}
                      \end{array}\right.
\ee
\be\label{eq:b}
(b)\quad a_{W_2}^b=\left\{\begin{array}{l}
                       a_{W_2}^M\quad \mbox{if}\quad |a_{W_2}^M|<|a_{W_2}^m|\\
                       a_{W_2}^m\quad \mbox{otherwise}
                      \end{array}\right.
\ee
In  Tab.~\ref{tab:aW_2Mm} we list some results for the case $z=0.8$ and 
$M_1=1$~TeV. A choice made between cases (a) and (b) above  
influences also the other fermion couplings, for example the $f\bar f Z_2$ vertices.  In particular,
we get for the above combination of $z$ and $M_1$ that (also assuming $a_{W_1}=0.13$)
\be\label{eq:aabb}
\frac{V_{f\bar fZ_2}^b}{V_{f\bar fZ_2}^a}\sim \frac{a_{W_2}^b}{a_{W_2}^a}=0.94,
\ee 
\be\label{eq:aabb1}
\frac{\Gamma_{W_2}^b}{\Gamma_{W_2}^a}\sim\frac{\Gamma_{Z_2}^b}{\Gamma_{Z_2}^a}\sim0.97.
\ee
\begin{table}[!htb]
\begin{center}
\begin{tabular}{||c||c|c|c||}
\hline \hline
$a_{W_1}$&$a_{W_2}^a$ & $a_{W_2}^b$& $a_{W_2}^b/a_{W_2}^a$ \\
\hline 
\hline
0.025&$-0.022$&$-0.006$&0.27\\
0.00& 0.035& 0.0018&0.51\\
$-0.02$&0.067&0.051&0.76\\
\hline\hline
\end{tabular}
\end{center}
\caption{Values for $a_{W_2}$ for three choices of $a_{W_1}$ considering both cases $(a)$ and $(b)$ defined in eqs. (\ref{eq:a}) and (\ref{eq:b}), respectively, for $z=0.8$ and  $M_1=1$~TeV.}
\label{tab:aW_2Mm}
\end{table}
One could presume that the differences between choice (a) and (b) induced onto 
the CS are negligible, especially when differences in masses and widths are rather small, as seen in eqs.~(\ref{eq:aabb})--(\ref{eq:aabb1}).  
While this may be true in some cases, it is not so generally. Regarding the case of processes (\ref{eq:processDY}) and (\ref{eq:processWW}), this assumption
has been verified to be correct in Refs.~\cite{Accomando:2010ir,Accomando:2011xi,Accomando:2012yg} (for the final choice of cuts used therein and imported here as well).
For process (\ref{eq:process}) studied here, this is true for the {\it C2} cuts (within typical uncertanties of CS measurements) but not the {\it Cc} ones, thereby further motivating the choice of the former
as default set for our 4S parameter scans. In fact, 
assuming again $z=0.8$ and $M_1=1$~TeV while taking, e.g., 
 $a_{W_1}=0.13$, we get (using {\it Cc} cuts) CS$_{a}=64$~fb and CS$_{b}=88$~fb and
(using {\it C2} cuts)  CS$_a=1.8$~fb and 
CS$_{b}=1.6$~fb. So it is clear that, on the one hand,
for a given mass spectrum there is a large range available for the CS and,
on the other hand,  even if the mass is experimentally measured, there remains an
uncertainty on the couplings. In essence, the numbers above already hint at the beneficial effects 
of the {\it C2} cuts, with respect to the {\it Cc} ones, as the differences in CS between the choices
in eqs.~(\ref{eq:a}) and (\ref{eq:b}) are much less in the former case than in the latter. 
We will further quantify this effect later on in Subsect.~\ref{MassSpetr}. We now proceed though to
compare channel (\ref{eq:process}) to processes (\ref{eq:processDY}) and (\ref{eq:processWW}) as
discovery modes of the 4S model. 

\subsection{Exclusion and Discovery}

We start this subsection by comparing the twin processes of mixed and charged di-boson production, i.e.,
 (\ref{eq:process}) and (\ref{eq:processWW}), respectively, e.g., at 14 TeV. At inclusive level, as we can  see from Tab.~\ref{tab:ahch}, the main difference between the former (denoted by $WZ$) and the latter
(denoted by $WW$) is a significantly different CS, notably due to a residual SM contribution entering
process (\ref{eq:process}) with resonant dynamics, which is instead absent in process (\ref{eq:processWW}). However, the
significance is better for the $WW$ case than for the $WZ$ one, primarily due to the aforementioned higher 
background induced by the SM in the latter case with respect to the former. Notice that, as
previously, 
the background is the SM yield whilst the signal is the difference between the full 4S result and
the SM itself. Similar results are obtained also at 7 and 8 TeV. 
Finally, such a pattern generally persists over the entire parameter space of the 4S
model. If one instead adopts the {\it C2} cuts, the $WW$ and $WZ$ rates become more compatible and the significances of the latter improve, to the
expense of lower signal rates. Hence, it makes sense to adopt the tighter set of cuts as the default one in the parameter scan, further considering that they are designed to exalt the resonant contributions of the new gauge bosons states
(unlike the looser set, which reveals a stronger sensitivity to interference effects), hence in tune with standard
experimental approaches.
\begin{table}[!t]
\begin{center}
\begin{tabular}{||c|c|c|c|c|c|c||}
\hline \hline
 $M_1$ (TeV)&0.5&0.75&1&1.5&1.7&2\\
\hline \hline
$WW$& 9.6 & 11.8 & 13.4 & 11.8 & 9.9 & 4.9 \\
\hline
$WZ$ ({\it Cc})&57.0/33.4&89.2/64.6& 89.0/63.8& 60.8/41.4& 56.8/36.4& 32.0/51.2\\
\hline
$WZ$ ({\it C2})&1.70/1.42&1.42/1.20&1.80/1.56&1.82/1.68&1.50/1.36&1.08/0.98\\
\hline\hline
\end{tabular}
\end{center}
\caption{Signal CS (in fb) defined as the total 4S rate minus the SM one at the LHC with $\sqrt s=14$ TeV 
considering both di-boson processes discussed in the text, 
for various values of $M_1$ and $z$ fixed to $0.8$, with the choice
of maximal allowed value for $a_{W_1}$. For the $WZ$ process the two values reported here
correspond to the solution (a)/(b)  in eq.~(\ref{eq:a})/(\ref{eq:b}), for both the {\it Cc} 
and {\it C2} cuts. The value for the SM 
is 1.55(5006/1.40)~fb for the $WW(WZ[{\it Cc/C2}])$ channel. For both signatures, the di-lepton one emerging
from $WW$ and the tri-lepton one stemming from $WZ$, we have taken different flavors only, $e\mu$, 
for consistency.
}
\label{tab:ahch}
\end{table}

\begin{figure}[!t]
\begin{center}
\unitlength1.0cm
\begin{picture}(7,4)
\put(-5.6,-4){\epsfig{file=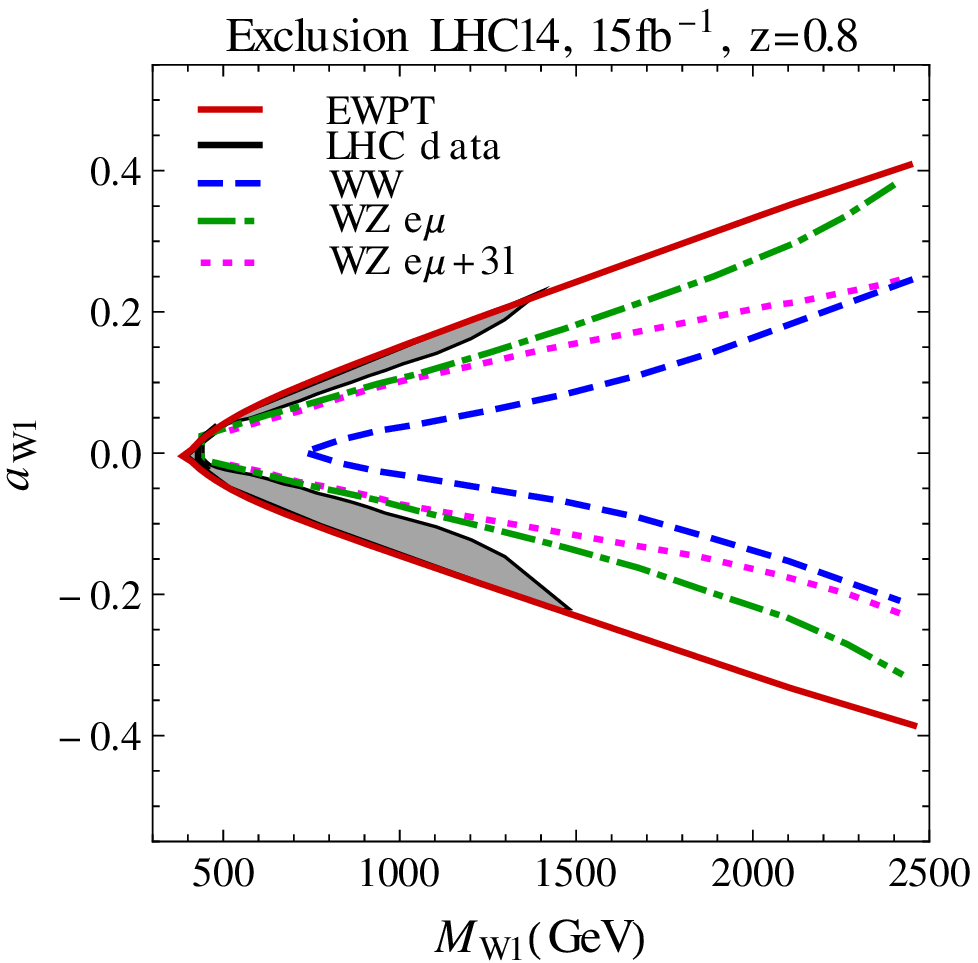,width=7.5cm}}
\put(3.5,-4){\epsfig{file=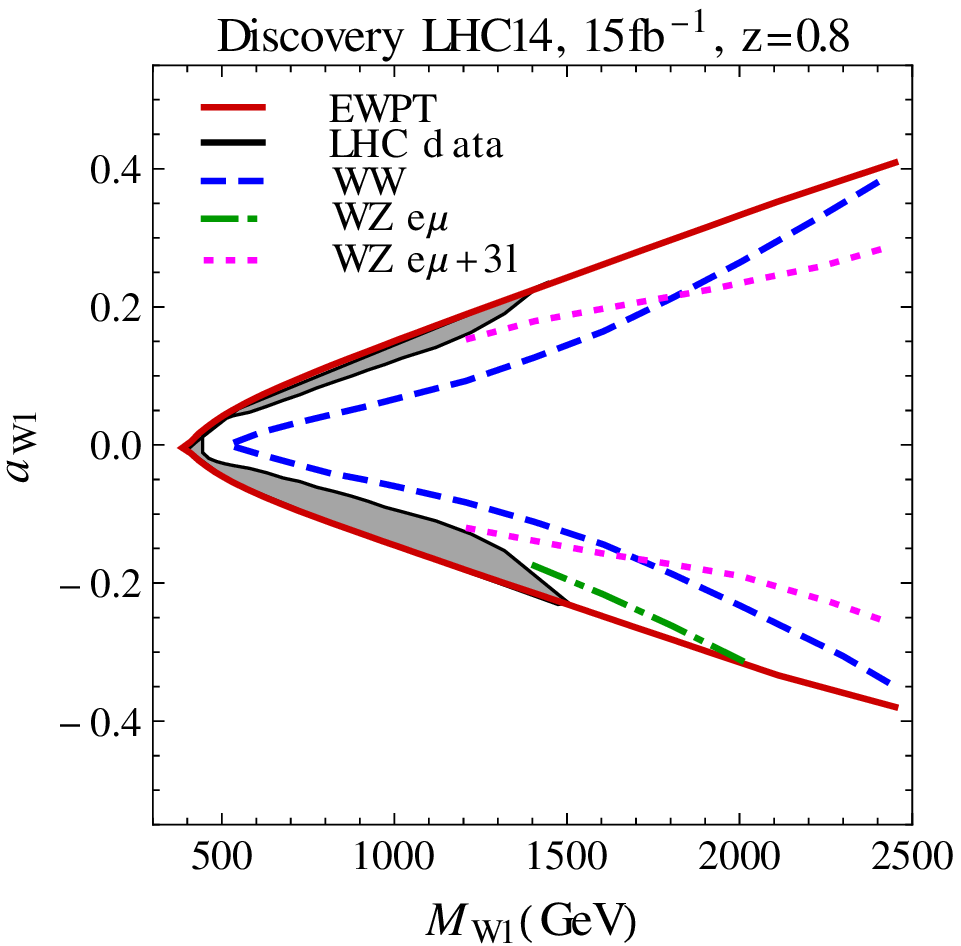,width=7.5cm}}
\end{picture}
\end{center}
\vskip 3.5cm
\caption{
Left(Right): Contrasting the $WZ$ and $WW$ yields in terms of 
$2\sigma$ exclusion($5\sigma$ discovery) contours over the plane ($M_{W_1}$, $a_{W_1}$) still allowed by EWPTs and direct searches via the DY modes for  the indicated luminosities. We are  using here the choice $(a)$ for $a_{W_2}$.  Here, we consider both the $e\mu$ and $3l$ final state in case of the $WZ$ process.
Also notice that the same detector efficiency for  leptons and missing transverse energy as extracted from data (see
Ref.~\cite{Accomando:2012yg} for further details) was used 
in all cases.
}
\label{fig:E+D_LHC14_C2}
\end{figure}

In the remainder of this section, we summarize the exclusion and discovery potential of the 14~TeV LHC
with respect to processes (\ref{eq:process})--(\ref{eq:processWW}). For the last two channels, we borrow 
some of the results obtained
in Ref.~\cite{Accomando:2012yg}. Herein, for illustration purposes, we consider again the value of $z=0.8$.
Assuming {\it C2} cuts and $e\mu$ final states for processes (\ref{eq:process})--(\ref{eq:processWW}),  
the CS decreases very quickly for $a_{W_1}$ going to 0, consequently, there are only very small excluded regions through $WZ$ in the $(M_{W_1},a_{W_1})$ plane. In particular, we get that the $WZ$ exclusion limits are 
always less stringent than those from $WW$. For what concerns discovery,  the results are quite similar.  This
is illustrated in Fig.~\ref{fig:E+D_LHC14_C2}, where  
we show the exclusion and discovery limits at the LHC with 14 TeV after
15~fb$^{-1}$ for processes (\ref{eq:process}) and (\ref{eq:processWW}), plotted against the regions still allowed by
EWPTs and direct searches via the processes in (\ref{eq:processDY}), the latter updated to the latest CMS and ATLAS
analyses based on 7 TeV data after 5 fb$^{-1}$, mapped over the $(M_{W_1},a_{W_1})$ plane compliant with
unitarity requirements. 
The above conclusions change though if, instead of solely relying on the $e\mu$ final state in the case of the $WZ$ process, we also include the $3l$ case
\footnote{Notice that di-lepton final states with identical flavors are of no use for the $WW$ process, as they are burdened by an overwhelming 
SM background induced by $ZZ$ events, with one $Z$ boson decaying invisibly. }. The same figure in fact also points out that process (\ref{eq:process})
equals or overtakes channel (\ref{eq:processWW}) (and in turn, also the modes in (\ref{eq:processDY}), see Ref. \cite{Accomando:2012yg}), for exclusion and discovery
purposes, respectively, at large gauge boson masses, yet still compatible with unitarity limits. Beyond these, for larger boson masses, because of kinematical reasons, the exclusion and discovery regions 
will eventualy close.

Finally, notice that we have used here the choice (a) for $a_{W_2}$, though results
are qualitatively the same for the case (b). In fact, even assuming 7 or 8 TeV for the LHC (and standard accrued luminosities
of 5 and 15 fb$^{-1}$, respectively), the overall pattern remains unchanged though quantitatively the scope of the $WZ$ mode is 
reduced in comparison to that of the $WW$ and DY ones, owing to the fact that high gauge boson masses are more unattainable at reduced energies and
luminosities. 

\subsection{Mass Spectroscopy}
\label{MassSpetr}

\begin{figure}[!htbp]
\begin{center}
\unitlength1.0cm
\begin{picture}(7,4)
\put(-5.6,-4){\epsfig{file=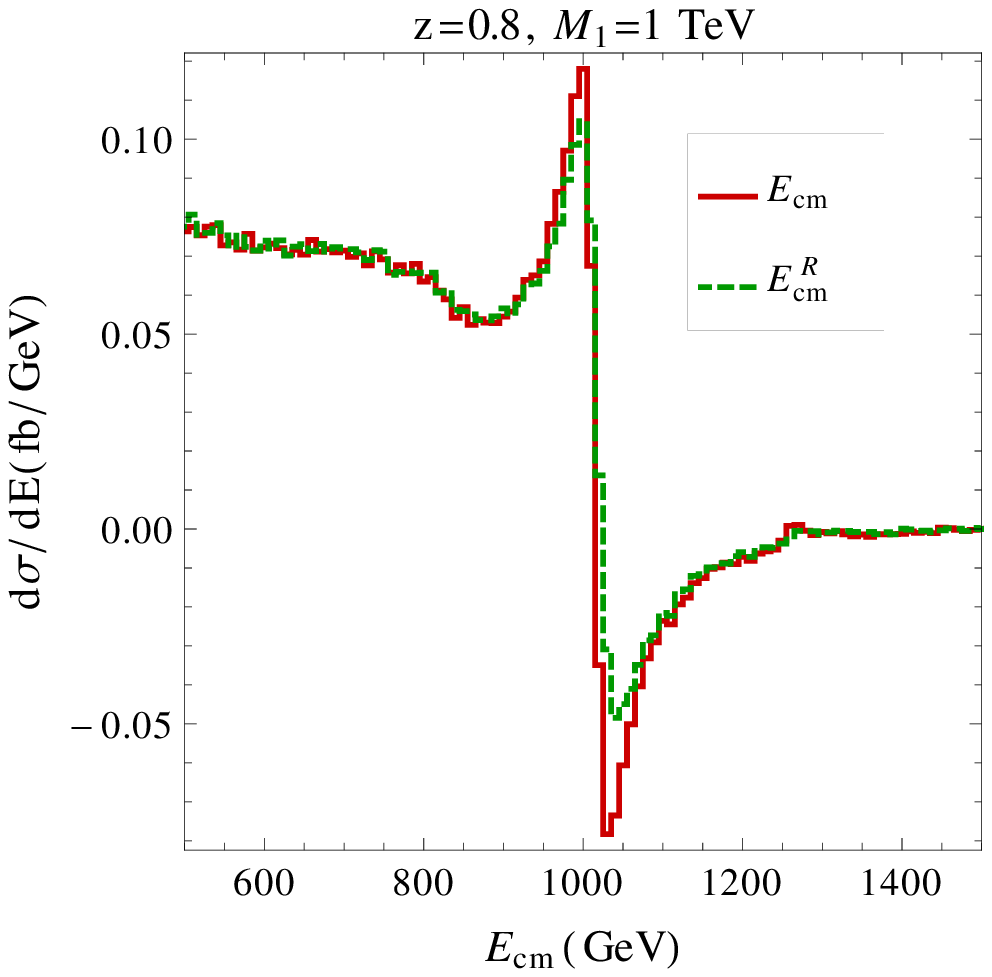,width=7.5cm}}
\put(3.5,-4){\epsfig{file=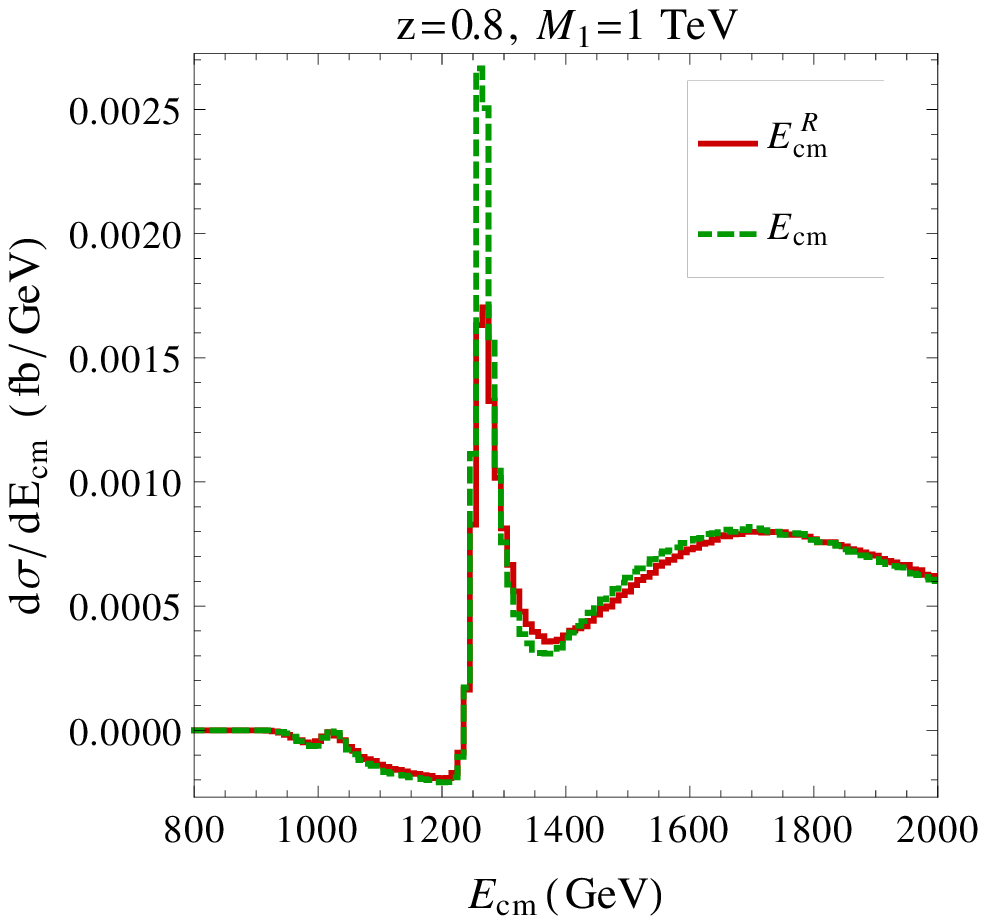,width=7.95cm}}
\end{picture}
\end{center}
\vskip 3.5cm
\caption{Differential distributions of the true ($E_{\rm cm}$) and reconstructed ($E^R_{\rm cm}$)
Center-of-Mass (CM) energy at partonic level for the signal, defined as the difference between the full 4S result
and the SM one, for a benchmark point in parameter space. In particular, we choose $z=0.8$, $M_1$=1~TeV and $a_{W_1}=0.13$. Left(Right): using the \textit{Cc}(\textit{C2}) cuts. 
}
\label{fig:Ecm}
\end{figure}

In this part of the paper, we intend to show that mixed di-boson production at the LHC 
(again, we take $\sqrt s=14$ TeV for illustration purposes) can act
as an effective means to extract the entire mass spectrum of the gauge bosons of the 4S model as well as to 
strongly constrain the size of their couplings. In order to accomplish this, it is crucial the fact that
process (\ref{eq:process}) affords one with the possibility to reconstruct the missing transverse momentum of the neutrino. In particular, in order to do so, 
we use the algorithm of Ref.~\cite{Bach:2011jy}, which reproduces a (partonic) invariant mass
($\sqrt{\hat s}=E_{\rm cm}$) distribution  that agrees with the true distribution very well: see Fig.~\ref{fig:Ecm} \footnote{Hereafter, we choose to use a 10 GeV bin for the differential distributions due to the fact that the experimental resolution (in mass and transverse momentum) is of order 1\% at 1--2~TeV for the electron, while for the muon the rate is more like 5\%.}. Incidentally, the left-hand plot here highlights the $W_1$ contribution, emerging
from the interference with the SM, while the right-hand plot singles out the $W_2$ contribution, stemming 
from the 4S resonance. In principle,
one should use all the three values of the intervening charged gauge boson masses (i.e.,
$W$, $W_1$ and $W_2$) to reconstruct the $\nu$ momentum. However, here we use only the SM  
mass, $M_W$, under the assumption that the corresponding resonance always gives a very sizable contribution
amongst the three. In fact, this represents an effective choice, for
both sets of cuts adopted throughout, i.e., {\it Cc} and {\it C2} \footnote{In contrast, note that the more kinematically constrained structure of DY processes does not allow one to apply the same algorithm in the case of the charged current channel in (\ref{eq:processDY}), so that here the extraction of the masses and
couplings of the 4S charged gauge bosons is only possible through a transverse mass \cite{Accomando:2011xi}, notoriously less sensitive to the actual resonant mass than an invariant mass distribution.}.

Again, the benchmark chosen here
is representative of a situation occurring generally over the 4S parameter space. 

\begin{figure}[!htbp]
\begin{center}
\unitlength1.0cm
\begin{picture}(7,10)
\put(-4.3,2.7){\epsfig{file=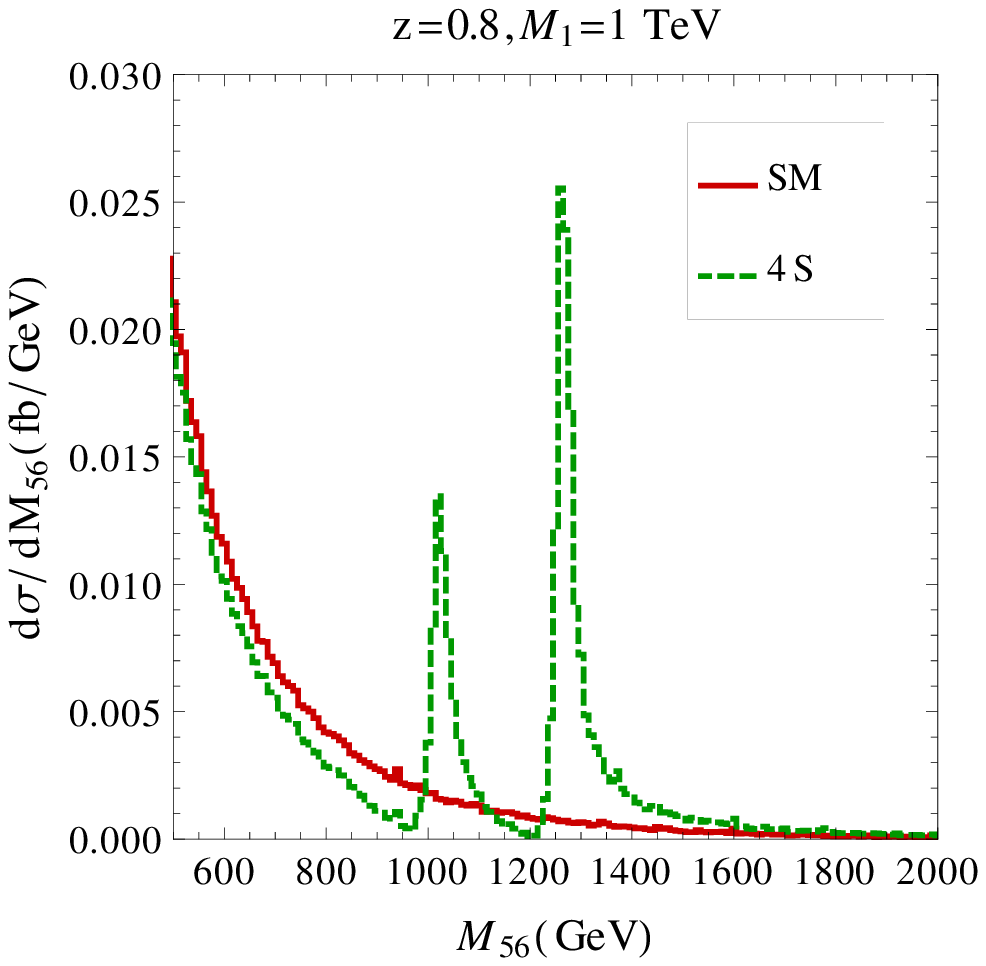,width=7.5cm}}
\put(3.5,2.7){\epsfig{file=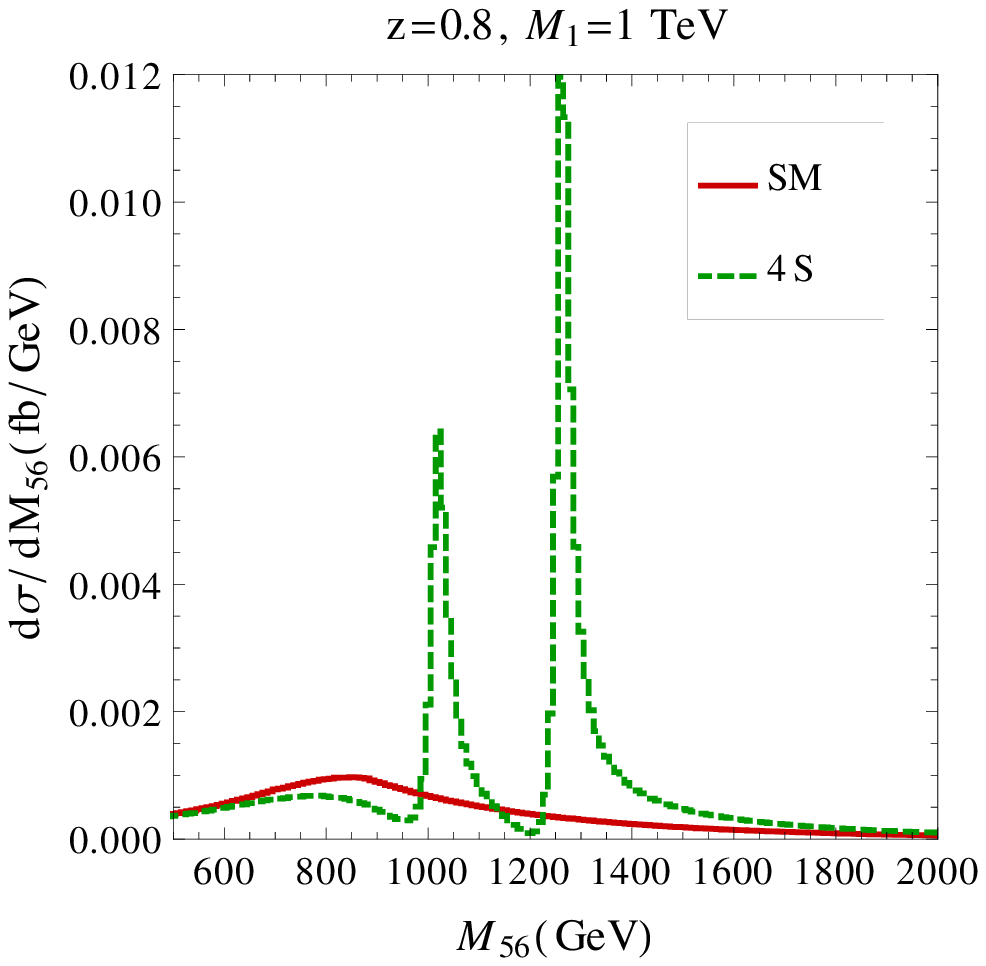,width=7.5cm}}
\put(-4.3,-5){\epsfig{file=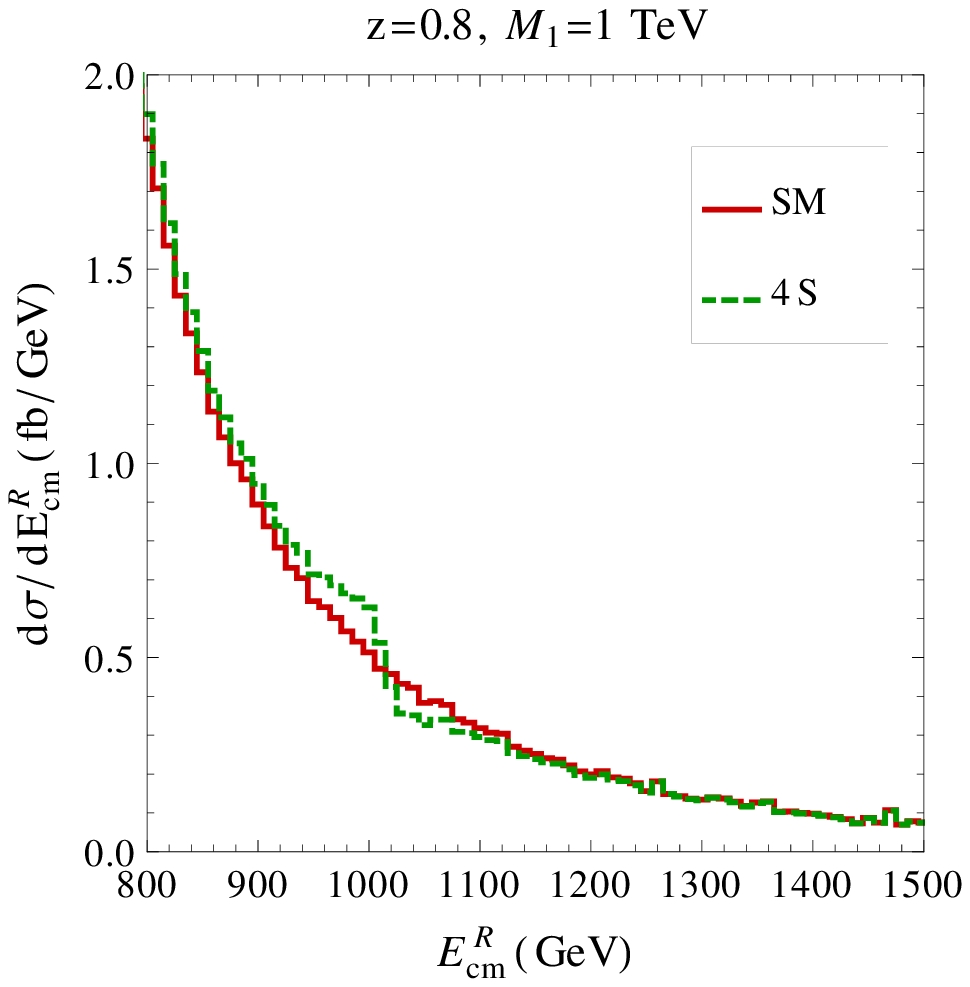,width=7.5cm}}
\put(3.5,-5){\epsfig{file=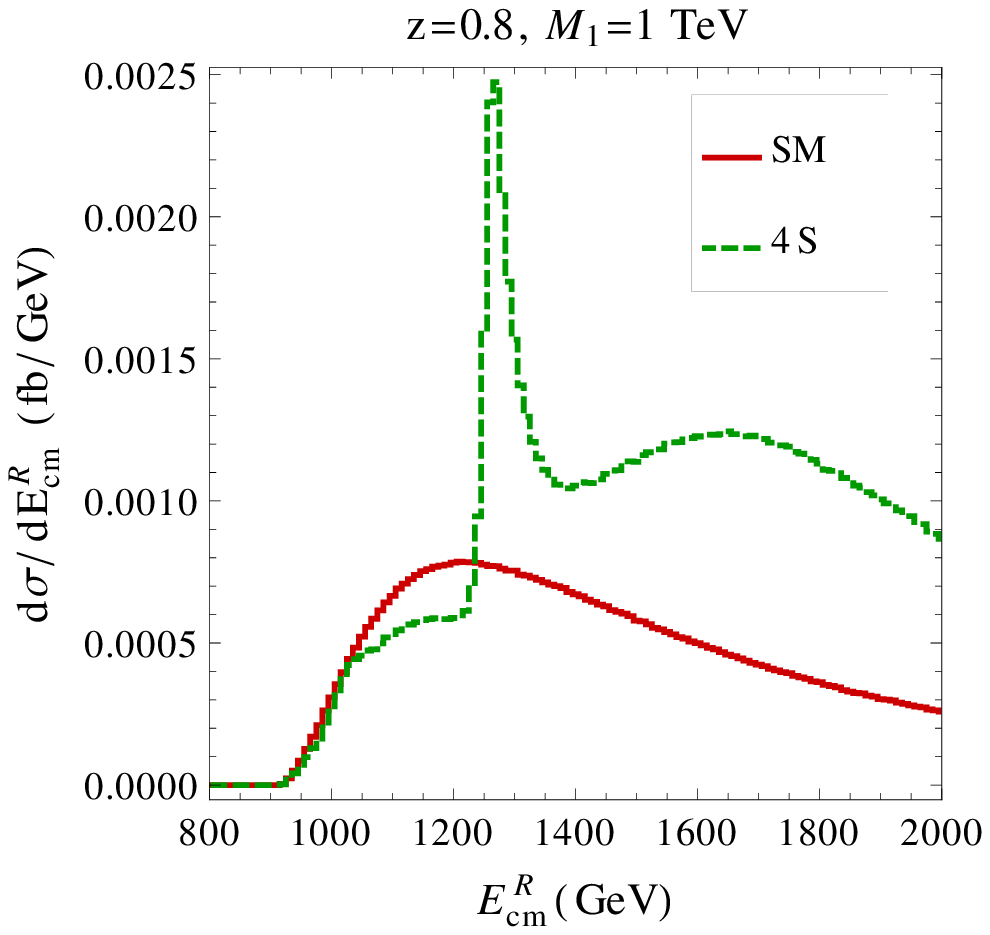,width=7.5cm}}
\end{picture}
\end{center}
\vskip 4.cm \caption{
Differential distributions of (top) the invariant mass of the di-lepton pair coming from the neutral gauge boson ($M_{56}$) and of (bottom) the
 reconstructed CM energy at partonic level ($E^R_{\rm cm}$) for the  full 4S model
and for the SM, for a benchmark point in parameter space. In particular, we choose $z=0.8$, $M_1$=1~TeV and $a_{W_1}=0.13$. Left(Right): using the \textit{Cc}(\textit{C2}) cuts.}
\label{fig:distr_C1_M56_M456}
\end{figure}

In Fig.~\ref{fig:distr_C1_M56_M456} we show the differential 
distributions in $M_{56}$ and  $E_{\rm cm}^R$ (the `reconstructed' $E_{\rm cm})$,  
 for a benchmark configuration of the 4S model ($z=0.8$, $M_1$=1~TeV, $a_{W_1}=0.13$). These 
two spectra are the most sensitive ones to the $Z_{1,2}$ and $W_{1,2}$ masses, respectively. 
The combination of the two set of cuts is necessary in order to extract the gauge boson mass spectrum
of the 4S model. On the one hand, the
{\it Cc} cuts can extract both neutral gauge boson resonances (top-left plot). For these in fact the {\it C2} set merely acts in the direction of further suppressing the SM background (notice the disappearance of
the low invariant mass tail in the top-right frame). On the other hand, the {\it C2} set of cuts becomes necessary if one wants to  study the properties of all the heavier resonances, as exemplified by comparing in this figure the
bottom-left (where only the $W_1$ resonance is barely visible) to the bottom-right 
                   (where the $W_2$ resonance clearly stands out) plot. 
In Tab.~\ref{Tab:banch} we show the masses and widths of the heavy gauge bosons for the aforementioned 4S benchmark point, so that one can easily identify the origin of the shapes seen in Fig.~\ref{fig:distr_C1_M56_M456}. 
\begin{table}[!htb]
\begin{center}
\begin{tabular}{||c|c|c|c||}
\hline \hline
 $Z$ &$M,\Gamma$ (GeV)&$W$&$M,\Gamma$ (GeV)\\
\hline \hline
1&   1012, 33.8 & 1 & 1008, 33.3 \\
\hline
2& 1256, 27.2 & 2 & 1255, 26.7  \\
\hline\hline
\end{tabular}
\end{center}
\caption{Numerical values of for masses and widths of the heavy gauge boson resonances for 
the benchmark point $z=0.8$, $M_1$=1~TeV and $a_{W_1}=0.13$ of the 4S model.}
\label{Tab:banch}
\end{table}

\begin{figure}[!t]
\begin{center}
\unitlength1.0cm
\begin{picture}(7,4)
\put(-5.6,-4){\epsfig{file=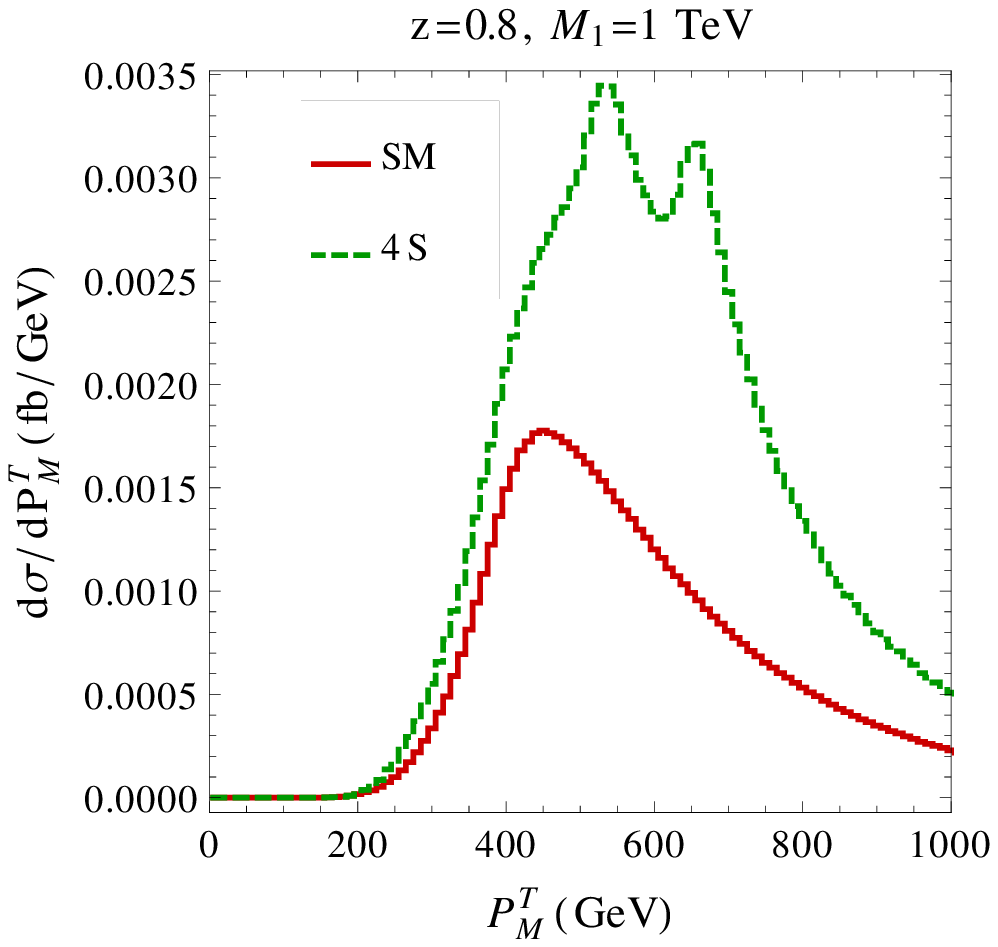,width=7.5cm}}
\put(3.5,-4){\epsfig{file=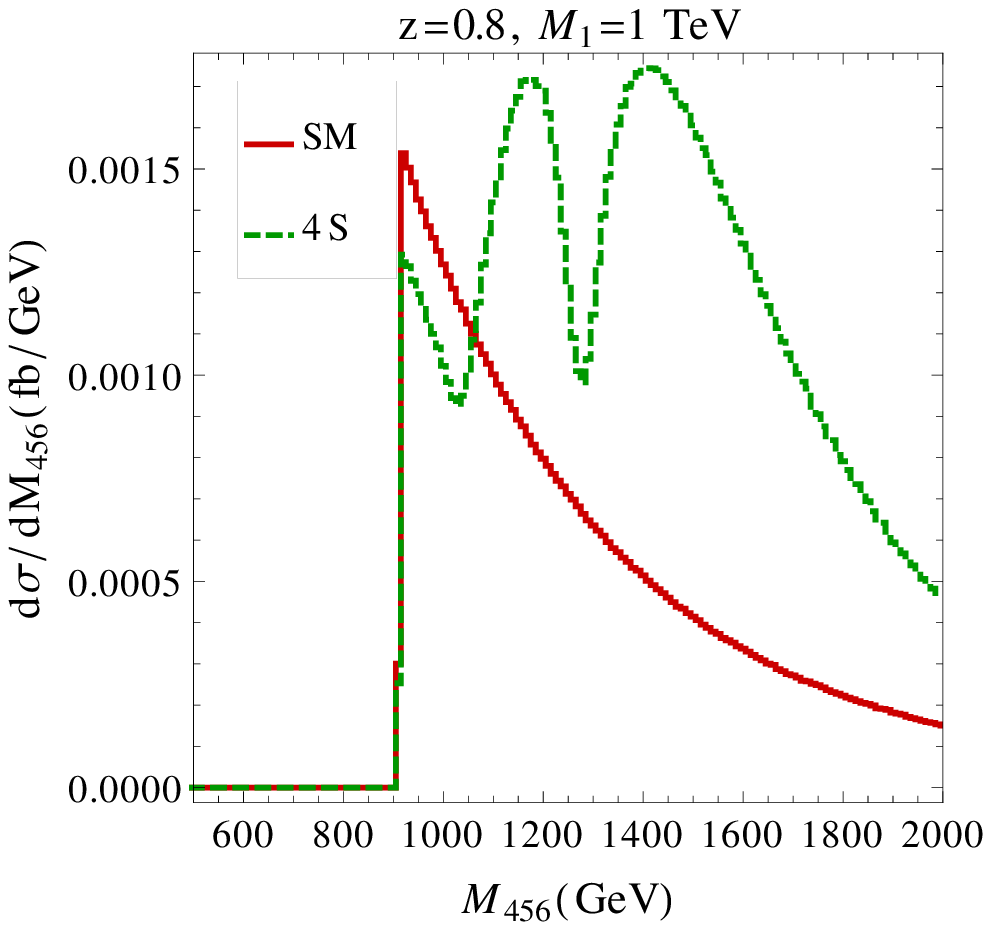,width=7.5cm}}
\end{picture}
\end{center}
\vskip 4cm
\caption{
Differential distributions of (left) the transverse momentum of the di-lepton pair coming from the neutral gauge boson ($P^T_{M}$) and of (right) the invariant mass of the three leptons ($M_{456}$) for the  full 4S model and for the SM one, for a benchmark point in parameter space. In particular, we choose $z=0.8$, $M_1$=1~TeV and $a_{W_1}=0.13$. Here, we are using \textit{C2} cuts.}
\label{fig:C2}
\end{figure}

Other kinematic variables which are efficient to establish the heavy gauge boson resonances of the 4S model  are
$P^T_{M}$ and $M_{456}$, so long that the {\it C2} cuts are used. This is exemplified in Fig.~\ref{fig:C2}. Here, one can appreciate the $Z_1$ and $Z_2$ resonances stemming in the first 
variable and the $W_1$ and $W_2$ ones emerging from the second observables. 
One subtlety to observe here though is that in the former case the terms responsible for the effect
are the squares of the corresponding 4S diagrams (the peaks in $P^T_M$ are exactly at $M_{Z_1}/2$
and $M_{Z_2}/2$, left plot) whereas in the latter case the relevant contributions are due to the interference of the corresponding 4S graphs with the SM ones (one can appreciate that are the dips
and not the peaks which correspond to the $W_1$ and $W_2$ masses, right plot). Again, we
illustrate such a phenomenology for the case of our usual benchmark scenario, though we can
confirm that it remains the same over the entire parameter space of the 4S model.

\begin{figure}[!t]
\begin{center}
\vspace{-.8cm}
\unitlength1.0cm
\begin{picture}(7,10)
\put(-4.3,2.7){\epsfig{file=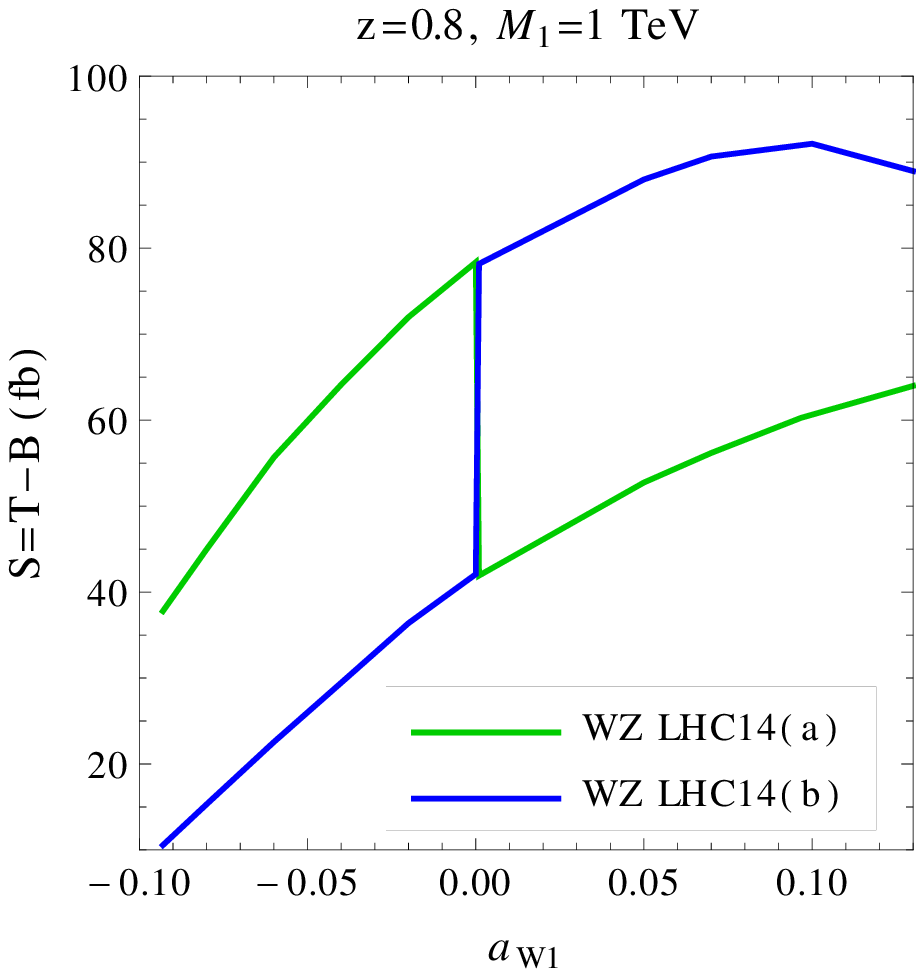,width=7.3cm}}
\put(3.5,2.7){\epsfig{file=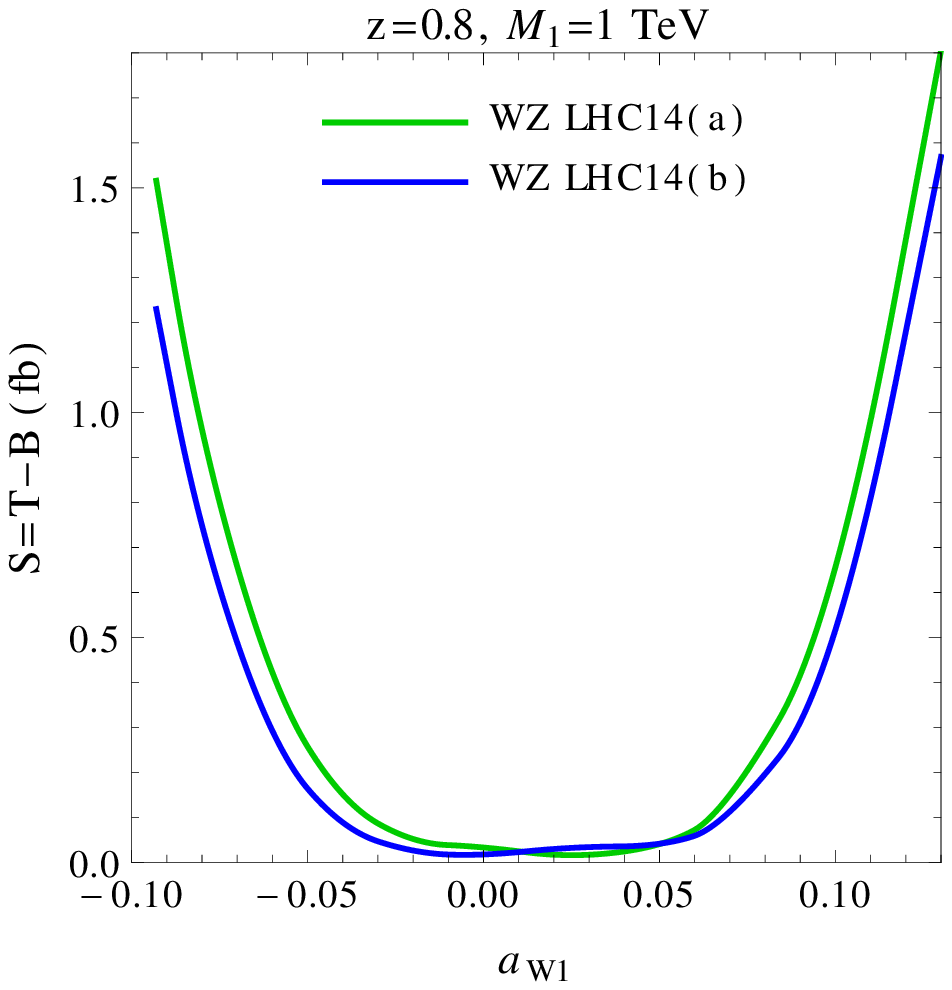,width=7.2cm}}
\put(-4.3,-5){\epsfig{file=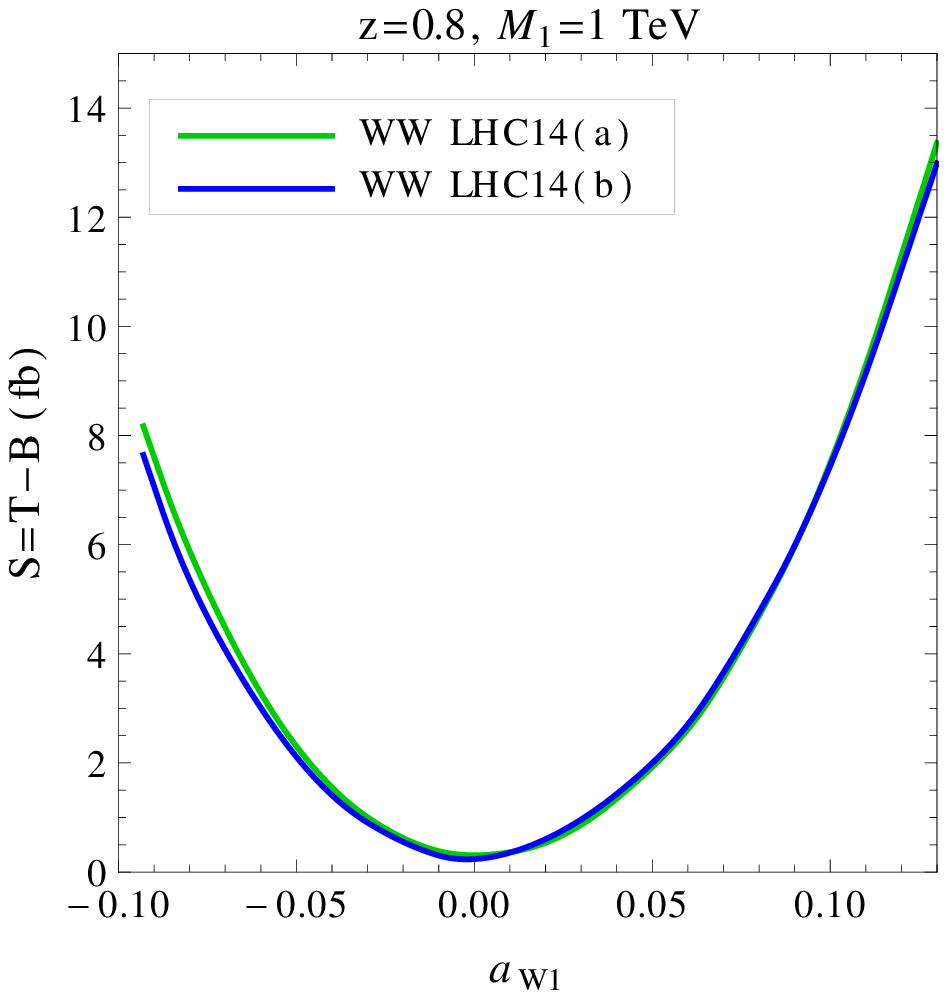,width=7.5cm}}
\put(3.5,-5){\epsfig{file=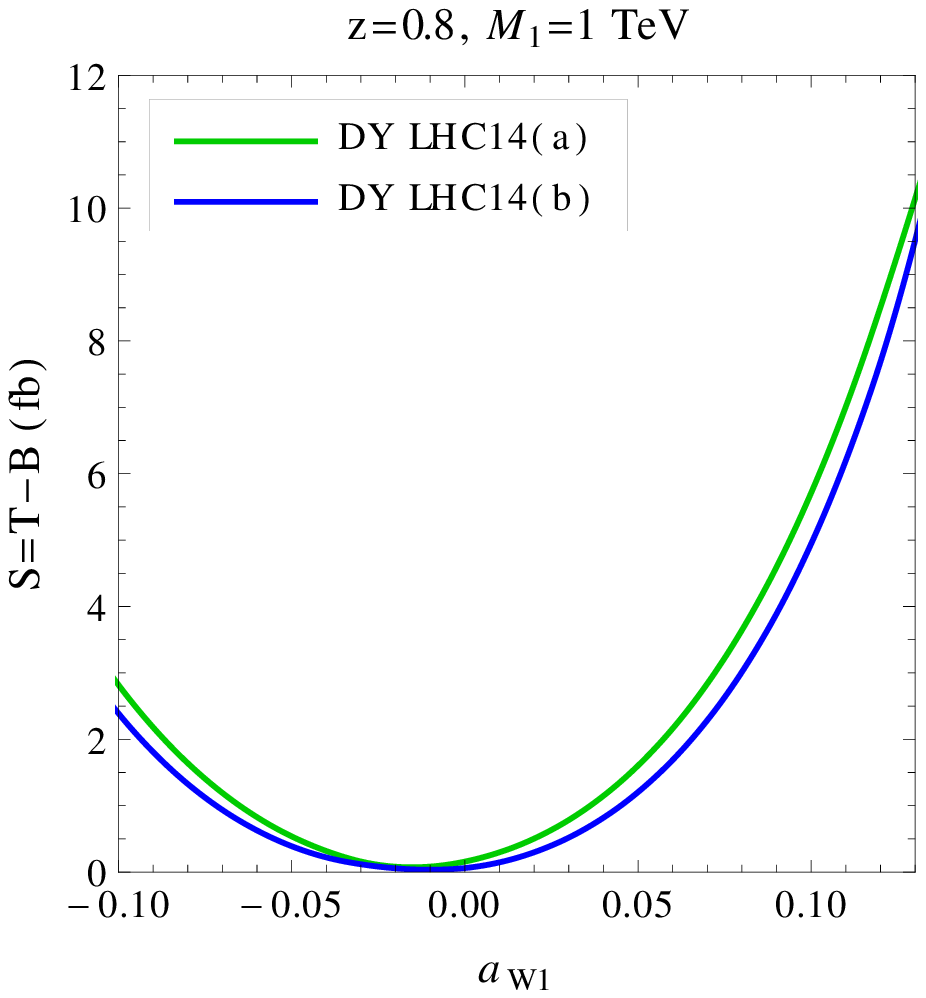,width=7.7cm}}
\end{picture}
\end{center}
\vskip 4cm \caption{Correlations between the CS for the $WZ$ signal process, defined as
the difference between the full 4S result and the SM one,  and the
coupling $a_{W_1}$ for the benchmark point $z=0.8$ and 
$M_1=1$~TeV. Top-left(right): using {\it Cc}({\it C2}) cuts.
Bottom-left(right): the same for $WW$(DY) using the {\it So}(standard) cuts introduced in \cite{Accomando:2012yg}(\cite{Accomando:2010ir,Accomando:2011xi}). 
The labels (a) and (b) refer to the solutions in eqs.~(\ref{eq:a}) and (\ref{eq:b}), respectively. 
These values refers to the $e\mu$ process.}
\label{fig:CS_Cc_mM_z08-1-2Tev}
\end{figure}

Furthermore, the lineshapes of the distributions used to extract the gauge resonances (with the {\it C2} cuts) of the 4S model are also
largely insensitive to the choice between the (a) and (b) solutions in eqs.~(\ref{eq:a}) and (\ref{eq:b}), respectively.  
In fact, another very interesting byproduct of the {\it C2} cuts is related to the possibility of strictly 
constraining the model couplings. In fact, as shown in a 
previous subsection, once all the masses of the model are extracted
(thereby $z$ and $M_1$ are known), two free parameters still
remain, $a_{W_1}$ and $a_{W_2}$. As we can see from Fig.~\ref{fig:CS_Cc_mM_z08-1-2Tev} (top-left plot), if the {\it Cc} set of cuts is adopted,
the interval over which, e.g., $a_{W_1}$ can be constrained is very large. In fact, it corresponds 
to the spacings between the points on the blue and green curves intercepting an horizontal line given
by a measured CS. Instead, if we consider the {\it C2} cuts, for a
fixed CS, we have only two very small allowed regions for 
$a_{W_1}$, see Fig.~\ref{fig:CS_Cc_mM_z08-1-2Tev} (top-right plot)
\footnote{This however occurs at the cost of a much reduced cross section, hence of a larger statistical error on the corresponding experimental measurement, so that a compromise
between the {\it Cc} and {\it C2} cuts may have to be found.}.
The same can be said for the $WW$ and DY
channels, see Fig.~\ref{fig:CS_Cc_mM_z08-1-2Tev} (bottom-left and bottom--right plots, respectively).
This is due to the fact that the {\it C2} cuts are very sensitive to $a_{W_1}$ 
(but not to $a_{W_2}$) so that, once we have a 
precise measurement of $a_{W_1}$, we can re-use the sample defined by {\it Cc} cuts (that are very sensitive to $a_{W_2}$  but not to $a_{W_1}$)  
in order to extract the appropriate value of $a_{W_2}$. Further notice that, despite, for consistency with the $WW$ case studied in Ref.~\cite{Accomando:2012yg},
we have limited the exercise here for the $WZ$ process to the case of the $e\mu$ flavor combination only, this can equally be done for the $3l$ one. 
Finally,  the same procedure
can be applied to any point on the 4S parameter space, although we have illustrated it here for our usual
benchmark point. 

In closing then, we should conclude that a judicious use of the two sets of cuts introduced can
enable one to extract efficiently both the mass and coupling spectrum of the 4S model. However, the spectrum reconstruction
described here requires rather large luminosities, so that, whilst it is certainly applicable to the LHC at 14 TeV, it remains 
of limited scope at 7 and 8 TeV.

\section{Conclusions and Outlooks}
\label{sec:summa}
In summary, mixed di-boson production (via $WZ$ diagrams) in a 4S model supplemented by a composite Higgs state revealed itself  as an important LHC channel in order to test such a scenario of EWSB. On the one hand, it crucially
contributes to the discovery potential of the LHC over new regions of the parameter space, if compared to the scope of 
the twin charged di-boson (via $WW$ topologies)  mode and DY processes, specifically, for high gauge boson masses. 
On the other hand, thanks to the fact that the source of missing
(transverse) energy in the case of the $WZ$ channel is due to only one neutrino (unlike the case of
$WW$ mediation, where it is due to two neutrinos), one can reconstruct fully the final state kinematics, hence
extracting all the intervening resonances, in turn implying a complete knowledge of the gauge boson spectrum of the
4S model. This can be achieved after the sequential implementation, firstly, of looser cuts 
merely emulating detector acceptance (that however
extract the typical mass scale of the lowest lying new resonances)   and, secondly, of a tighter selection
exploiting such kinematical information (that enables an effective mass spectroscopy). Such conclusions are
generally valid independently of the LHC setup, though they become quantitatively most relevant at higher energies and luminosities.

Finally, we should emphasise that a byproduct of our analysis was the realization that the aforementioned
looser cuts are primarily sensitive to interference effects (between the genuine 4S contributions and the SM ones)
whereas the tighter ones enhance instead resonance effects (primarily of the new heavy gauge bosons). This therefore calls, in the same
spirit as in Ref.~\cite{Accomando:2011eu} in the DY case, for revisiting the scope of the di-boson channels (both charged
and mixed) at the LHC, under assumptions different from those routinely made by the LHC experiments (of new resonance dominance).
Also, one other aspect so far unexplored of processes (\ref{eq:process}) and (\ref{eq:processWW}) which does not pertain to the 
channels in (\ref{eq:processDY}) is the dependence of the CS on tri-linear gauge couplings {(i.e., the three gauge boson
vertices, see Appendix~\ref{app1})}, which will also constitute the subject of another
publication.

\section*{Acknowledgments}
We all thank Elena Accomando for discussions.
SM is financed in part through the NExT Institute. LF thanks Fondazione Della Riccia for financial support. 
 The work of SDC and  DD  is partly supported by
the Italian Ministero dell'Istruzione, dell'Universit\`a
e della Ricerca (MIUR) under the COFIN programme (PRIN 2008).

\appendix
\section{Tri-linear gauge boson vertices}
\label{app1}
Here we list the tri-linear gauge boson vertices of the 4S model. The full heavy gauge Lagrangian is presented in~\cite{Accomando:2008dm,Accomando:2008jh,Accomando:2010ir}, in particular the general tri-linear gauge boson vertex has the following structure:
\be
T_{PQR}=(P_{\mu\nu}^+Q^{+\mu}R^{3\nu}-Q_{\mu\nu}^-P^{+\mu}R^{3\nu}+R_{\mu\nu}^3P^{+\mu}Q^{-\nu}+h.c.)
\ee
and the Lagrangian can be written as
\be\label{eq:3l}
\mathcal{L}=i\sum_{PQR}a_{PQR}T_{PQR}
\ee
where $P,Q=W,W_1,W_2$ and $R=\gamma,Z,Z_1,Z_2$. In Tab.~\ref{tab:3l} we give the analytical expressions for $a_{PQR}$ at the leading order in $x=\sqrt{2}scM_Z/[M_1\sqrt{(1-z^2)}]$, with $s$ and $c$ the sine and cosine of the Weinberg angle defined as in~\cite{Accomando:2008jh}. Clearly $a_{PQR}=a_{QPR}$ and $a_{W_iW_j\gamma}=e\delta_{ij}$ with $e$ the electric charge.

\begin{table}[!htb]
\begin{center}
\begin{tabular}{||c||c|c|c||}
\hline \hline
 &$Z$&$Z_1$&$Z_2$\\
\hline \hline
$WW$&$ {\frac{e{ c}}{{ s}}}$&$ {-\frac{ex}{2\sqrt{2}{ s}}(1-z^4)}$&$ {\mathcal{O}(x^3)}$\\
\hline
$WW_1$&$ {-\frac{ex}{2\sqrt{2}{ s}{ c}}(1-z^4)}$&$ {\frac{e}{2{ s}}}$&$ {\frac{ez^2}{2{ s}}}$\\
\hline
$WW_2$&$\mathcal{O}(x^3) $&$ {\frac{ez^2}{2{ s}}}$&$ {\frac{e}{2{ s}}}$\\
\hline
$W_1W_1$&$ {\frac{e}{2{ s}{ c}}({ c}^2-{ s}^2)}$&$ {\frac{e}{\sqrt{2}{ s}x}}$&$ {-\frac{exz^2}{2\sqrt{2}{ s}{ c}^2}\frac{2{ c}^2-z^2({ c}^2-{ s}^2)}{1-z^2}}$\\
\hline
$W_1W_2$&$ {\frac{ez^2}{2{ s}{ c}}}$&$ {\frac{exz^2}{2\sqrt{2}{ s}{ c}^2}\frac{{ c}^2(z^2-2)+{ s}^2}{1-z^2}}$&$ {\frac{e}{\sqrt{2}{ s}x}}$\\
\hline
$W_2W_2$&$ {\frac{e}{2{ s}{ c}}({ c}^2-{ s}^2)}$&$ {\frac{e}{\sqrt{2}{ s}x}}$&$ {\frac{exz^4}{2\sqrt{2}{ s}{ c}^2}\frac{4{ c}^2-1}{1-z^2}}$\\
\hline\hline
\end{tabular}
\end{center}
\caption{Analytical expressions for the tri-linear couplings at leading order in $x$, as described in eq.~(\ref{eq:3l}).
}
\label{tab:3l}
\end{table}

\hyphenation{Post-Script Sprin-ger}


\end{document}